\documentstyle[aps,prb,epsf]{revtex} 

\newcommand{\hcc}{H_{\mbox{\scriptsize cc}}}
\newcommand{\hcd}{H_{\mbox{\scriptsize cd}}}
\newcommand{\hdc}{H_{\mbox{\scriptsize dc}}}
\newcommand{\hdd}{H_{\mbox{\scriptsize dd}}}
\newcommand{\bge}{\begin{equation}}
\newcommand{\ee}{\end{equation}}
\newcommand{\eqm}{\rightleftharpoons}

\begin{document}

\title{
\begin{flushleft}
{\footnotesize OU-THP-9613S}\\
{\footnotesize BU-CCS-960301}\\
{\footnotesize {\it J. Amer. Chem. Soc. {\bf 118} (1996) 10719-10724}}\\[0.3in]
\end{flushleft}
{\bf Simulation of Self-Reproducing Micelles using a
Lattice-Gas Automaton}}
\author{
{\bf Peter V. Coveney} \\
Schlumberger Cambridge Research,\\
High Cross, Madingley Road, Cambridge CB3 0EL, U.K.\\
{\bf Andrew N. Emerton} \\
Theoretical Physics, Department of Physics, University of Oxford \\ 
1 Keble Road, Oxford OX1 3NP, U.K.\\
{\bf Bruce M. Boghosian},\\
Center for Computational Science, Boston University, \\
3 Cummington Street, Boston,Massachusetts 02215, U.S.A. \\
}
\date{April, 1996}
\maketitle

\begin{abstract}
  We simulate self-reproducing micellar systems using a recently
  introduced lattice-gas automaton~\cite{bib:bce}. This dynamical model
  correctly describes the equilibrium and non-equilibrium properties of
  mixtures of oil, water and surfactants.  The simulations reported here
  mimic the experiments of Luisi {\em et al.}~\cite{bib:blln} in which
  caprylate micelles are formed by alkaline hydrolysis of immiscible
  ethyl caprylate ester. As in the laboratory experiments, we find an
  extended induction period during which the concentration of micelles
  remains small; thereafter the ester is consumed very rapidly with
  concomitant production of micelles.
\end{abstract}

\section{Introduction} 
\label{sec:intro}

This article reports on simulations of autopoietic self-reproducing
micellar systems that we have carried out using our recently introduced
lattice-gas model for self-assembling amphiphilic
systems~\cite{bib:bce}. A self-reproducing system is one in which the
process leading to population growth, here of amphiphile and micelles,
occurs as a result of the parent structures themselves. Additionally, a
structure which is self-bounded and is able to self-generate due to
reactions which take place at or within that boundary, meets the
criteria of {\em autopoiesis}, which certain authors have advocated as a
minimal definition of life~\cite{bib:vmu,bib:fcl}.

Our model simulates the behavior of various experimental systems that
have been investigated in recent years. Experiments on micelles that can
catalyse their own reproduction were first described in and around
$1990$-$1991$~\cite{bib:bwlle,bib:bll,bib:bwlll}. In these, micelles
were present in the initial reaction mixture -- that is, the system was
presented with the bounded structures required for autocatalysis. Then
in $1992$ Bachmann, Luisi and Lang~\cite{bib:blln} reported on a
two-phase system in which autocatalytic micelles were formed from
amphiphiles that were themselves generated from a hydrolysis reaction
within the aqueous solution. The system contained an aqueous sodium
hydroxide solution and an ester, ethyl caprylate, which is itself
immiscible with water. During the reaction, the system is well stirred
in order to enhance the mixing of the two liquids. Hydrolysis of the
ester initially produces the amphiphile monomer, sodium caprylate, which
is known to form micelles in water~\cite{bib:ff}. This process occurs at
a ``slow'' background reaction rate, largely confined to the ester-water
interface. However, as soon as sufficient caprylate is present in the
system for aggregation into micelles to take place (which occurs at and
beyond the critical micelle concentration, in the terminology of
equilibrium thermodynamics), a sudden increase in the reaction rate is
observed. Bachmann {\it et al.} interpreted this as being due to a
micellar catalytic process: When a micelle is formed, its hydrophobic
interior causes it to rapidly bind oily ethyl caprylate molecules,
dramatically enhancing the solubility of the previously immiscible
liquid. The rate of hydrolysis is suddenly increased because the ester
is now dispersed throughout the aqueous medium and its probability of
encountering water-soluble hydroxide ions is vastly greater. Moreover,
new amphiphile is produced by a chemical reaction that takes place at or
within the domain boundary of the parent micelles, thereby qualifying as
an instance of autopoietic self-reproduction. This amphiphile is then
able to form more micelles, and the hydrolysis reaction accelerates. 
The latter reaction process is the ``fast''
micellar catalysis step, the rate of micelle production increasing as
the micelle concentration increases.
The reaction is completed when there is no longer any ethyl caprylate
left in the system. Upon lowering the pH of the system,
Bachmann {\em et al.} found that the product micelles could be
reversibly converted into vesicles formed from closed caprylate bilayer
membranes.  These authors argued that the properties of such systems may
have relevance to chemical mechanisms associated with the origins of
life on Earth~\cite{bib:blln}.

In outline, the physicochemical steps occurring can be written down as
rate processes:
\begin{eqnarray}
EC & \rightarrow & C, \nonumber \\
nC & \rightarrow & C_{n}, \label{CMC_mod}  \\
C_{n} + EC & \rightarrow & C_{n} + C. \nonumber
\end{eqnarray}
where $EC$ denotes ethyl caprylate, $C$ the caprylate anion and $C_{n}$
a caprylate micelle ($n$ being the number of caprylate monomers per
micelle, estimated~\cite{bib:blln} to be approximately 63). The third
step here represents the micellar-catalysed hydrolysis of ester.

Although the qualitative explanation of the Luisi experiments proffered
here is simple enough to comprehend, producing a quantitative model of
the phenomenon is quite a different matter. The difficulties stem from
the need to describe the dynamics of micelle formation correctly. Only
recently has a detailed nonlinear dynamical model of micelle formation
kinetics been proposed and analysed~\cite{bib:cw}, with particular
reference to Luisi's system (see also~\cite{bib:bc}). The approach is
based on a generalisation of the Becker-D\"{o}ring scheme describing the
kinetics of first-order phase transitions~\cite{bib:bd}, in which
micellar clusters grow (shrink) through the stepwise addition (removal)
of individual surfactant monomers. The general kinetic process in the
Becker-D\"{o}ring theory can be written as
\begin{equation}
C_{r} + C \;\stackrel{a_{r}, b_{r+1}}{\eqm} \; C_{r+1}, \nonumber
\end{equation}
where $C_{r}$ is a cluster comprising $r$ monomers and $r$ ranges from
one to infinity. Here $a_{r}$ and $b_{r+1}$ are the rate coefficients
for the forward and reverse steps respectively. The reaction rates are
given by the law of mass action, and lead to an infinite system of
coupled nonlinear differential equations which may be combined with the
processes shown in the scheme~(\ref{CMC_mod})~\cite{bib:cw}.
 
At a more microscopic level, one would like to be able to describe
micellar dynamics accurately. This is one aspect of the wider issue of
the challenge of modelling self-assembling amphiphilic systems. Although
in principle one might wish to use molecular dynamics to simulate the
interactions in these systems in full molecular detail, in practice such
an approach is computationally overbearing and will remain so for the
foreseeable future given the state of today's technology. Moreover, some
of the most important situations demanding attention involve the
coupling of self-assembly with hydrodynamics~\cite{bib:jmsk}. Their
complexity is the
result of the interplay between microscopic interactions and mesoscale
hydrodynamic effects. Accordingly, we have developed a lattice-gas model
of such systems which is capable of addressing these issues in a
computationally tractable manner.

The purpose of the present paper is to show how our model may be applied
to simulate the dynamics of self-replicating micelles. Even at
equilibrium, such structures are highly dynamic, rapidly exchanging
monomers with the surrounding medium on timescales which are typically
of the order of micro- to nano-seconds. This makes these processes very
difficult to capture using molecular dynamics (MD), where the
integration time-steps are on the order of femtoseconds. By contrast,
lattice gas automata are based on a discrete time-stepping procedure
performed on a spatially discrete lattice: The interactions are confined
to collisions at the vertices, and the particles are then advected to
neighboring sites in time steps that are on the order of the mean free
time, several orders of magnitude larger than those in MD.

The formation of micelles from surfactant monomers is an important part
of the simulations described here. In addition, however, in the
self-reproducing micelle experiments a chemical reaction is involved
which converts the oily ester into surfactant molecules. Again, it is
a strength of cellular
automata that they enable chemical reactions to be included easily by
means of simple collisional transition rules; examples include the
Belousov-Zhabotinski reaction~\cite{bib:MH}, the Schl\"{o}gl
model~\cite{bib:bmb}, and cement hydration~\cite{bib:bcgks}. We shall
find that our suitably modified lattice-gas automaton model displays the
same behaviour as that shown in the real system of self-reproducing
micelles. It therefore bridges the divide between the two approaches to
the investigation of autopoietic structures advocated by
Varela~\cite{bib:fv}: virtual (computational) and physical (chemical
synthesis). Cellular automata were originally proposed as providing a
virtual model of autopoiesis by Varela {\em et al.}~\cite{bib:vmu}; it
is therefore fitting that we are able to connect the logical and the
physical through a finite-state automaton simulation of the physical
approach. Indeed, we hope that our results may be seen as being of wider
significance than purely as simulations of phenomena in the natural
world. For in recent years, there has been a renewed interest in the
concept of ``artificial life,'' the study of life as it could be (in
whatever shape or form it may be found or made on Earth or elsewhere),
rather than as it is known today on Earth \cite{bib:ch}. According to
this view, neither actual nor possible life is determined by the matter
of which it is composed. For life is a {\em process}, and it is the {\em
  form} of this process, not the matter, that is the essence of life. As
von Neumann originally sought to demonstrate, one can ignore the
physical medium and concentrate on the {\em logic} governing this
process~\cite{bib:jvn}. In principle, one can thus achieve the same
logic in another clothing, totally distinct from the carbon-based form
of life we know.

\section{Description of the Model}
\label{sec:maa}

Our lattice-gas model is a microscopic dynamical system which gives the
correct mesoscopic and macroscopic behaviour of mixtures of oil, water
and surfactant. The model is based on the two-fluid immiscible lattice
gas of Rothman and Keller~\cite{bib:rk}, which we have reformulated
using a microscopic particulate decription to permit the inclusion of
amphiphile.  The model exhibits the commonly formed equilibrium
microemulsion phases, including droplets, bicontinua and
lamellae~\cite{bib:bce}. Moreover, the model conserves momentum as well
as the masses of the various species, and correctly simulates the fluid
dynamical and scaling behaviour during self-assembly of these
phases~\cite{bib:fhp,bib:ecb}.

In order to incorporate the most general form of interaction energy
(Hamiltonian) within our model system, we introduce a set of coupling
constants $\alpha, \mu, \epsilon, \zeta$, in terms of which the total
interaction energy can be written as~\cite{bib:bce}
\begin{equation}
\Delta H_{\mbox{\scriptsize int}}
       =   \alpha \Delta \hcc +
         \mu \Delta \hcd +
         \epsilon \Delta \hdc +
         \zeta \Delta \hdd.
\label{eq:tiw}
\end{equation}
The four terms on the right hand side correspond, respectively, to the
relative immiscibility of oil and water, the tendency of surrounding
surfactant to bend around oil or water droplets, the propensity of
surfactant dipoles to align across oil-water interfaces and the
contribution from pairwise interactions between surfactant molecules.

In the experiments of Bachmann {\it et al.}~\cite{bib:blln}, the caprylate
amphiphile produced has a tendency to form micelles in the
bulk water phase rather than predominantly congregating in monolayers at the
oil-water interface. Previous studies with our model~\cite{bib:bce}
enable us to simulate this property by choosing the following set of
coupling constants:
\bge
 \alpha = 1.0, \mu = 1.0, \epsilon = 3.0, \zeta = 0.5,
 \label{eq:dcc}
\ee
for our simulations.  There is another parameter present in our model
that must be specified, called $\beta$. This coefficient arises from the
stochastic, Monte-Carlo-like, nature of the selection of outgoing
collisional states at vertices within our lattice-gas automaton, and can
be thought of as an effective inverse temperature-like
parameter~\cite{bib:ecb}.

{}From the discussion of Luisi's self-reproducing micelle experiments
given in Sec.~\ref{sec:intro}, it is clear that the autocatalytic nature
of the reaction is related to the critical micelle concentration
(c.m.c.) of the caprylate anion.  Experimentally, this is the point when
the {\em equilibrium} concentration of surfactant monomers reaches a
level at which they associate into micelles. However, it is important to
recognise that the c.m.c. is not the location of a sharp phase
transition, but rather the concentration at which the equilibrium
fraction of monomers in micelles reaches some arbitrary value, usually
taken to be $0.5$. Indeed, the equilibrium fraction of monomers within
micelles is itself somewhat arbitrary, since there is no clear-cut
division between ``micelles'' and smaller (or, for that matter, larger)
clusters that are not regarded as micelles. The equilibrium fraction of
monomers in micelles varies very rapidly with monomer concentration
around the c.m.c., so that to an experimentalist it may look like a
phase transition. As in our work based on the generalised
Becker-D\"{o}ring equations~\cite{bib:cw}, our lattice-gas model
dispenses completely with the commonly made assumption of thermodynamic
equilibrium between monomers and monodisperse micelles, being a highly
dynamical, nonequilibrium model. Thus, we do not assume the
existence of a critical micelle concentration: Rather, the phenomenon is
itself an emergent property of our model.

At the mesoscopic level, our model is very useful for analysing the
kinetics of micellar growth. Indeed our model has already been shown to
exhibit a critical micelle concentration in the binary -- surfactant and
water (or oil) -- limit~\cite{bib:bce}. Using the set of coupling
constants as defined in Eq.~\ref{eq:dcc}, we reproduce this analysis in
order to obtain an estimate for the number of surfactant particles
required to be in the binary water-surfactant system for the c.m.c. to
be exceeded. The results can be found in Sec.~\ref{sec:res}.

Although an extension of our full microemulsion model to three
dimensions is currently underway~\cite{bib:bce3d}, in this letter we
shall be concerned only with the two-dimensional ($2D$) version. This is
sufficient to capture the salient properties of the micellar
self-production experiments; we expect the results in the $3D$ case to
be qualitatively similar.

In the basic simulation (as in the experiments) the system is initially
comprised of two immiscible fluids: The majority phase we identify with
the aqueous solution in the experimental set-up; the other, minority,
phase we take to be representative of the ethyl caprylate ester. We then
specify two chemical reaction mechanisms designed to represent: (i) the
``slow'' pseudo-first-order alkaline hydrolysis of caprylate by
hydroxide ions present within the aqueous medium producing surfactant,
and (ii) the ``fast'' micellar-catalysed hydrolysis during which
self-assembled micelles greatly enhance the solubility of ester within
the water phase and thus lead to an increased alkaline hydrolysis. As in
the actual experiments, the simulation reaches completion once all the
ester in the system has been hydrolysed.

Within our lattice-gas automaton, alkaline hydrolysis is described
algorithmically by the conversion of an ester (oil) particle into an
amphiphile particle if certain prescribed conditions are met; the system
is otherwise allowed to evolve in the normal way (that is, with
collisions conserving momenta and particle masses) and a record is kept
of both the total number of surfactant particles in the system and the
total number of micelles. (The precise manner in which ``micelles'' are
defined in these simulations is specified below.)  These data can then
be used to analyse our simulations in a manner similar to that used for
the experimental data generated by Bachmann {\it et
  al.}~\cite{bib:blln}.
 
In order to simulate the self-reproduction of micelles, we set up the
model as follows:

\begin{enumerate}

\item The $2D$ simulation domain with periodic boundary conditions in
  both dimensions is initialised with water and oil in a $6 : 4$ ratio,
  the two bulk fluid regions being entirely phase-separated in accord
  with the experimental starting conditions.
 
\item The inverse-temperature parameter $\beta$ was set to the value
  $0.137$. This value corresponds to the oil and water acting as
  immiscible fluids during the simulation (that is, below the so-called
  spinodal point at which the fluids become miscible), but
  where the natural fluctuations within the system (which are larger
  near the spinodal point) encourage invasion of the individual bulk,
  phase-separated fluid regions by particles of the opposite kind. This
  has a similar effect to continuous stirring in the experimental
  system, since both enhance the ester hydrolysis rate, in essence by
  increasing the interfacial area.

\item At each timestep and for every site ${\bf x}$ on the lattice which
  contains {\em at least one oil} particle, we perform the following
  analysis
 
\begin{itemize} 

\item ``Slow'' (pseudo-first-order) hydrolysis step: If five of the six
  nearest-neighbor sites are dominated by water particles, and less
  than two of these nearest-neighbor sites have amphiphile present
  (recall that there can be up to seven particles per site), then one of
  the oil particles on site ${\bf x}$ is converted into a surfactant
  particle with unit probability, prior to determining the outgoing
  state of the collision process from the look-up tables.  This
  simulates ester hydrolysis in an overwhelmingly aqueous environment.

\item ``Fast'' micellar catalysis step: If five or more of the six
  nearest-neighbor sites contain one or more surfactant particles and
  five or more of the six next-nearest-neighbor sites have water
  particles present, we define this structure as being equivalent to a
  micelle$^{*}$.
    \footnotetext{$^{*}$Note that in reality there are {\em twelve}
    next-nearest-neighbor sites on a hexagonal lattice. The six we
    specify here are those that lie along the directions defined by the
    six nearest-neighbor lattice directions from site ${\bf x}$: This
    defines points around an essentially circular structure (which would
    be spherical in 3D), is computationally simple to implement and
    suffices for our present purposes.}
Within this bounded structure a
  surfactant particle is created from an oil particle on site ${\bf x}$
  with probability one, prior to determining the outgoing state of the
  collision process from the look-up tables. If there is more than one
  oil particle at site ${\bf x}$ then the oil particle actually
  converted is chosen at random; this prevents any bias from developing
  in the velocity given to the newly created surfactant particles.
  Chemically speaking, this step simulates the hydrolysis of an ester
  molecule solubilised within a micelle, by hydroxide ions in the
  surrounding aqueous environment.

\item Otherwise the automaton is updated according to the usual
  collision rules~\cite{bib:bce}.

\end{itemize}

\item We keep track of the total number of surfactant particles present
  in the system as well as the number of micelles (as defined above)
  existing at each time step. Note that in addition to these
  ``oil-in-water'' type micelles, we also need to take into account
  similar nearest- and next-nearest-neighbor structures for which the
  central site is surfactant dominated. Otherwise the number of micelles
  present in the system would appear to drop to zero once all the oil in
  our system is converted to amphiphile.
 
\end{enumerate}

\section{Simulation Results}
\label{sec:res}

\begin{figure}
\begin{center}
\leavevmode
\hbox{
\epsfysize=2.8in
\epsffile{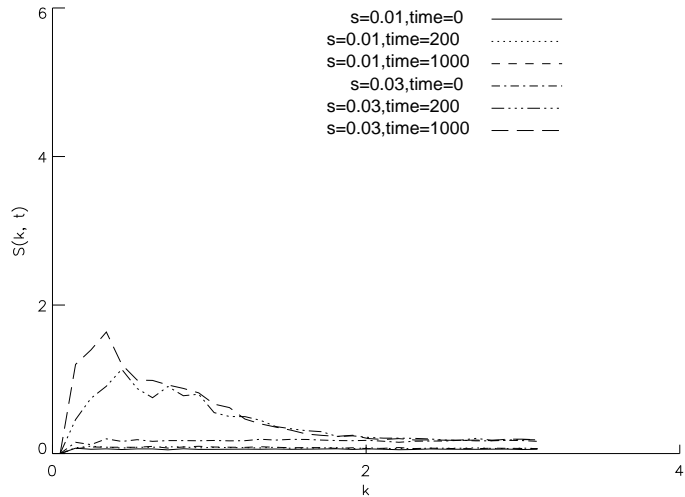}}
\end{center}
\caption{Temporal evolution of surfactant density structure function 
  S(k, t) for binary
  water and surfactant mixtures. The figure contains results for both
  $0.01$ and $0.03$ reduced density of surfactant.}
\label{fig:cmca} 
\end{figure}

\begin{figure}
\begin{center}
\leavevmode
\hbox{
\epsfysize=2.8in
\epsffile{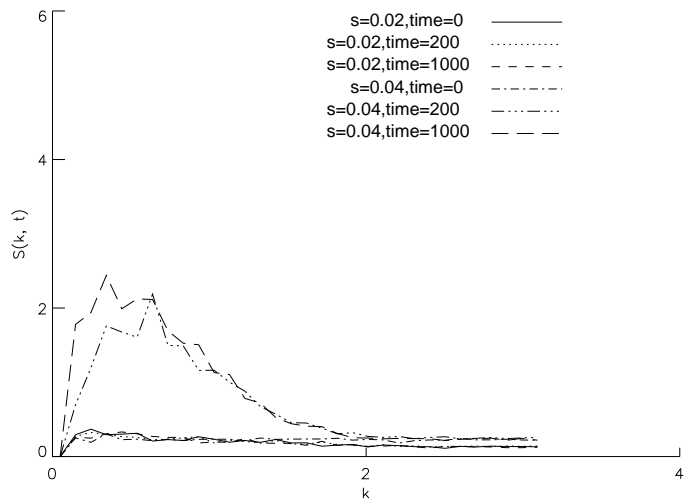}}
\end{center}
\caption{Temporal evolution of surfactant density structure function 
  S(k, t) for binary
  water and surfactant mixtures. The figure contains results for both
  $0.02$ and $0.04$ reduced density of surfactant.}
\label{fig:cmcb} 
\end{figure}

Initially we report on the analysis of the c.m.c. in a binary
water-surfactant limit of our model. Simulations are on a $64 \times 64$
lattice with the initial condition being a random distribution of the
water and surfactant particles; the density of surfactant in the system
is altered in each case. When sufficient amphiphile is present for the
monomers to form micelles, these impart characteristic structure to the
system which should be discernible in our simulations. To perform a
quantitative analysis we keep track of the circularly-averaged structure
functions $S(k,t)$ of the surfactant density~\cite{bib:bce}. The micelles
themselves are dynamic structures and are not necessarily very long
lived; in addition energetic considerations mean that individual
micelles do not grow without limit, hence we do not expect to see
evidence of micelles coalescing and growing in an unbounded manner. The
results of our simulations are shown in Figs.~\ref{fig:cmca} and
\ref{fig:cmcb}.

{}From Fig.~\ref{fig:cmca} it is clear that for a reduced density of
$0.01$ surfactant ($\approx 287$ surfactant particles) there is no
structure formation during the timescale of the run, so the amphiphile
exists as monomers. In contrast, for $0.03$ surfactant ($\approx 860$
surfactant particles) the figure shows structure formation early in the
simulation which is maintained and does not appear to grow without
limit, indicating that we have just reached the c.m.c. To further
clarify the approximate position of the c.m.c., Fig.~\ref{fig:cmcb}
shows the results for reduced densities of $0.02$ ($\approx 573$
surfactant particles) and $0.04$ ($\approx 1147$ surfactant particles)
surfactant.  The former of these is below and the latter above the
c.m.c. for this system. These results indicate that the number of
surfactant particles required to be in the system for the c.m.c. to be
exceeded lies between $573$ and $860$.

\begin{figure}
\begin{center}
\leavevmode
\hbox{
\epsfysize=2.8in
\epsffile{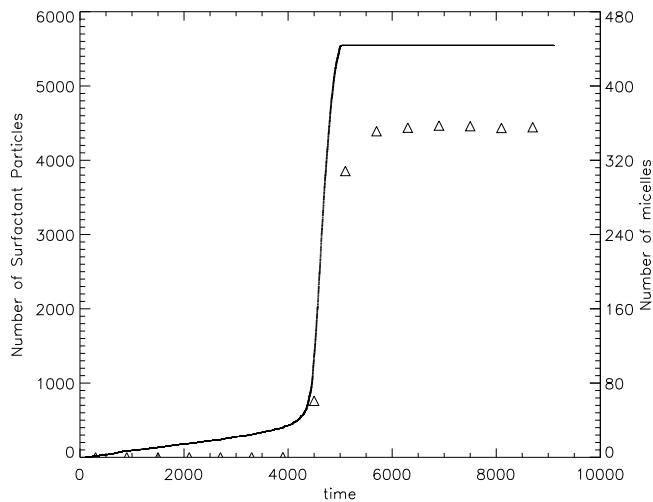}}
\end{center}
\caption{Amphiphile and micelle concentration as a function of time
  step in the lattice-gas automaton model. The triangles are for the
  micelles and correspond to the values indicated on the right hand
  vertical axis.}
\label{fig:owm} 
\end{figure}

\begin{figure}
\begin{center}
\leavevmode
\hbox{
\epsfysize=1.6in
\epsffile{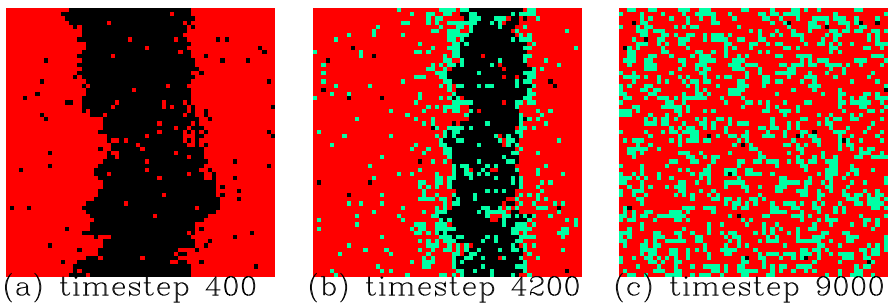}}
\end{center}
\caption{Three snapshots in time from the evolution of the lattice-gas
  automaton simulation. Ethyl caprylate molecules are black, water is
  red and anionic caprylate is green.} 
\label{fig:datowm} 
\end{figure}

The result of a typical simulation of self-reproducing micelles 
using the lattice-gas model described
in Sec.~\ref{sec:maa} is shown in Fig.~\ref{fig:owm}. The plot consists
of the temporal evolution (in dimensionless lattice-gas automaton time
steps) of the number of surfactant particles in the system as well as
the number of micelles on a lattice of size $64 \times 64$. The
comparison with the experimental results of Bachmann {\it et al.} is
good: We see a relatively long initial induction period during which
surfactant particles are produced by the ``slow,'' pseudo-first-order
hydrolysis step only; this is followed by an extremely rapid increase in
the number of surfactant particles in the system which occurs at exactly
the point at which micelles start to form. For, in accord with the 
estimate given above for the binary system, the amphiphile
concentration at this point ($\approx 600-700$ surfactant particles on
the lattice) is at the critical micelle concentration.
Figure~\ref{fig:datowm} shows a sequence of snap-shots from
the above described simulation at three different moments during the
simulation: (a) shows the situation shortly after the initial time, when
fluctuations have caused particles from the immiscible fluids to 
interpenetrate; (b) depicts the mixture at the onset of the rapid phase
in the reaction as amphiphile begins to be generated freely; and (c)
displays the homogeneous end state of the reaction, where all the oil
(ester) has been converted into amphiphile. 

\begin{figure}[tb]
\begin{center}
\leavevmode
\hbox{
\epsfysize=2.8in
\epsffile{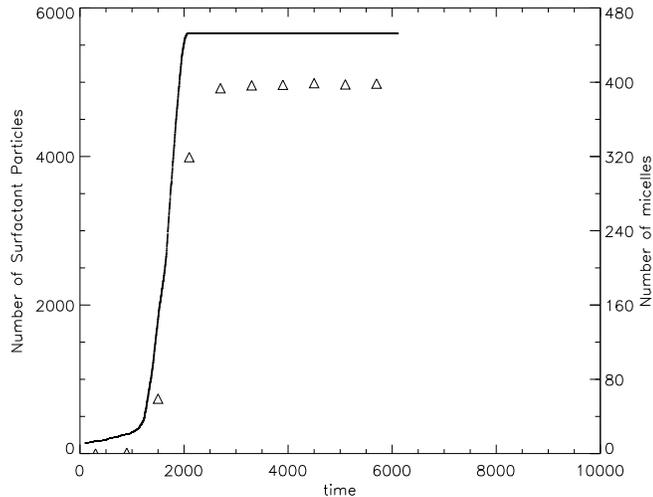}} 
\end{center}
\caption{Amphiphile and micelle concentration as a function of time
  step in the lattice-gas automaton model. The triangles are for the
  micelles and correspond to values on the right hand vertical axis.
  This case has surfactant in the system at time zero.}
\label{fig:owmko} 
\end{figure}

Fig.~\ref{fig:owmko} contains the result for an exactly analogous
simulation to that described above but now the system is given a
non-zero initial surfactant concentration. We use a reduced density of
$0.005$ which corresponds to about $150$ surfactant particles randomly
spread around the $64 \times 64$ lattice. The presence of these
particles should induce an earlier onset of the rapid growth phase since
the system should reach its critical micelle concentration more quickly.
In accord with the experimental results of Bachmann {\it et
  al.}~\cite{bib:blln}, this is exactly the behavior that we observe.

Note that in Fig.~\ref{fig:owmko}, the calculation of the number of
micelles present at each timestep is done with approximately 
$150$ extra surfactant particles in the system than is the case for
Fig.~\ref{fig:owm}. This accounts for the perceived discrepancy in the
number of micelles found in solution after the reaction has completed in
each case. The  system shown in Fig.~\ref{fig:owmko} is found to have
a slightly higher number of micelles
because of the extra amphiphile monomers initially present in the
solution. Again, by looking for the end of the ``slow'' hydrolysis step,
we can estimate the critical micelle concentration for this system; as
expected it is found to be approximately the same as in the previous
simulation and so is also in agreement with the prediction from the
binary water-surfactant mixture.

It must be remembered that these lattice-gas simulations are of a
microscopic, particulate nature, in contradistinction with our nonlinear
dynamical models~\cite{bib:cw}. They are therefore subject to
statistical fluctuations.  Accordingly, we have also performed a series
of simulations on progressively larger 2D lattices ($64 \times 64$, $128
\times 128$ and $256 \times 256$) in order to investigate the role of
these fluctuations. At the same time, we have ensured that, in all
cases, the simulations start from the {\em same} interfacial area to
volume ratio for the oily ester (in 2D, this is more correctly stated as
the interfacial length to area ratio), as this is recognised to be an
important parameter in the experimental set-up~\cite{bib:blln}.
(Experimentally, more rapid stirring leads to an enhanced
ester/aqueous-phase interfacial area and hence a faster overall rate of
reaction.) We carried out five such simulations for each size of
lattice, differing only in the precise specification of the initial
condition -- the distribution of initial particle velocities within the
bulk regions at timestep zero is randomised with a different seed for
each run, leading to different fluctuation effects in each case. The
results are summarised in Table I, which lists
statistics for the time 
taken for the ``slow'' induction period to come to an end and so also
marks the point at which the rapid reaction phase begins.

\begin{table} 
\vspace{0.1cm}
\centering
{
\renewcommand{\baselinestretch}{1}
\small\normalsize
\caption{{\bf Effect of lattice size on end of induction phase in
    lattice-gas automaton simulations of Luisi's experiments }} } 
\vspace{0.3cm}
\begin{tabular}{cccc}
\hline
\em{Lattice size}    &  \em{Average timestep}  
&  \em{Standard}  &   \em{Furthest deviation} \\
    &  \em{$t_0$ of ``fast'' phase} &  \em{deviation} & \em{from
$t_0$}\\ 
\hline
$64 \times 64$    &   3827    &    784.71   &  1193  \\
$128 \times 128$  &   3821    &    210.84   &   302  \\
$256 \times 256$  &   3835    &    144.52   &   221  \\
\hline
\end{tabular}
\label{tab:noise}
\end{table}

We find that the average timestep at which kick-off occurs is
essentially constant across these simulations, while deviations from
average behaviour are greatest for the smallest simulation box size and
least for the largest. This is exactly what one would expect: As the
number of particles in the system becomes larger, fluctuations are less
significant and the average behavior (of an ensemble of similar
systems) becomes ever closer to that observed experimentally.

\section{Conclusions}

Using our recently introduced lattice-gas model which describes mixtures
of surfactant, oil and water~\cite{bib:bce}, we have been able to
simulate the experimentally observed kinetics of Luisi's
self-reproducing micelles~\cite{bib:blln}. In the simulations, a
characteristic induction period is followed, at the onset of the model's
critical micelle concentration, by a very rapid production of surfactant
and micelles. The induction period is shortened by the initial addition
of surfactant, and by increasing temperature (which can also be taken to
correspond to higher stirring rates of the fluids inside the reaction
vessel). The length of the induction period is also sensitive to
fluctuations within the system; however, for ensembles of systems in
which the interfacial area to volume ratio of the ester initially
present is held constant, the ensemble average of the induction time is
essentially independent of system size.

Our simulations bring out very clearly one essential element in the
self-reproduction of caprylate micelles during the aqueous alkaline hydrolysis
of ethyl caprylate. This is the highly {\em dynamic} nature of the
entire process, including the reversible self-assembly and break-up of
the micelles themselves, which are also central features of our previous
Becker-D\"{o}ring model~\cite{bib:cw}. Moreover, ``micelles'' generally
persist as clusters of a range of sizes.  The self-reproducing micelle
experiment has been proposed as a paradigm of an autopoietic
process~\cite{bib:lv}, but it must be remembered that the micellar
structures are themselves evanescent -- they are continually breaking up
and reforming.  This is a property shared with living things, whose
overall forms may remain largely invariant but whose detailed molecular
constitutions are perpetually in a state of flux.

\section*{Acknowledgments}

PVC is indebted to Pier Luigi Luisi and Peter Walde for several
enlightening discussions. ANE thanks EPSRC and Schlumberger Cambridge
Research for funding his research. PVC and BMB thank NATO for partial
support for this project. BMB is supported in part by Phillips
Laboratory and by the United States Air Force Office of Scientific
Research under grant number F49620-95-1-0285.

\end{document}